\documentclass[twocolumn,showpacs,preprintnumbers,amsmath,amssymb,superscriptaddress]{revtex4}
\usepackage[bookmarks, pdftitle={ms_spinel_v12}, pdfauthor={Nair}, colorlinks=true, linkcolor=black, citecolor=blue, urlcolor=black]{hyperref}
\usepackage{amssymb}
\usepackage{amsmath}
\usepackage{graphicx}
\usepackage{dcolumn}
\usepackage{bm}
\begin{document}
%
%
%
%
\title
{Approaching the Ground State of Frustrated $A$-site Spinels: A Combined Magnetization and Polarized Neutron Scattering Study}
\author{Harikrishnan S. Nair}
\email{h.nair@fz-juelich.de}
\affiliation{J\"{u}lich Center for Neutron Sciences JCNS and Peter Gr\"{u}nberg Institute PGI, JARA-FIT, Forschungszentrum J\"{u}lich GmbH, 52425 J\"{u}lich, Germany}
\author{Zhendong Fu}
\affiliation{J\"{u}lich Center for Neutron Sciences JCNS, Outstation at MLZ, Forschungszentrum J\"{u}lich GmbH, Lichtenberg Stra{\ss}e 1, D-85747 Garching, M\"{u}nchen, Germany}
\author{J\"{o}rg Voigt}
\affiliation{J\"{u}lich Center for Neutron Sciences JCNS and Peter Gr\"{u}nberg Institute PGI, JARA-FIT, Forschungszentrum J\"{u}lich GmbH, 52425 J\"{u}lich, Germany}
\author{Yixi Su}
\affiliation{J\"{u}lich Center for Neutron Sciences JCNS, Outstation at MLZ, Forschungszentrum J\"{u}lich GmbH, Lichtenberg Stra{\ss}e 1, D-85747 Garching, M\"{u}nchen, Germany}
\author{Th. Br\"{u}ckel}
\affiliation{J\"{u}lich Center for Neutron Sciences JCNS and Peter Gr\"{u}nberg Institute PGI, JARA-FIT, Forschungszentrum J\"{u}lich GmbH, 52425 J\"{u}lich, Germany}
\affiliation{J\"{u}lich Center for Neutron Sciences JCNS, Outstation at MLZ, Forschungszentrum J\"{u}lich GmbH, Lichtenberg Stra{\ss}e 1, D-85747 Garching, M\"{u}nchen, Germany}
\date{\today}
\begin{abstract}
We re-investigate the magnetically frustrated, {\it diamond-lattice-antiferromagnet} spinels FeAl$_2$O$_4$ and MnAl$_2$O$_4$ using magnetization measurements
and diffuse scattering of polarized neutrons. 
In FeAl$_2$O$_4$, macroscopic measurements evidence a "cusp'' in zero field-cooled susceptibility around 13~K. Dynamic magnetic susceptibility and {\it memory effect} experiments provide results that do not conform with a canonical spin-glass scenario in this material. 
Through polarized neutron scattering studies, absence of long-range magnetic order down to 4~K is confirmed in FeAl$_2$O$_4$. By modeling the powder averaged differential magnetic neutron scattering cross-section, we estimate that the spin-spin correlations in this compound extend up to the third nearest-neighbour shell. 
The estimated value of the Land\'{e} $g$ factor points towards orbital contributions from Fe$^{2+}$. 
This is also supported by a Curie-Weiss analysis of the magnetic susceptibility. 
MnAl$_2$O$_4$, on the contrary, undergoes a magnetic phase transition into a long-range ordered state below $\approx$ 40~K, which is confirmed by macroscopic measurements and polarized neutron diffraction. 
However, the polarized neutron studies reveal the existence of prominent spin-fluctuations co-existing with long-range antiferromagnetic order. 
The magnetic diffuse intensity suggests a similar short range order as in FeAl$_2$O$_4$. Results of the present work supports the importance of spin-spin correlations in understanding magnetic response of frustrated magnets like $A$-site spinels which have predominant short-range spin correlations reminiscent of the "spin liquid'' state.
\end{abstract}
\maketitle
\section{Introduction}
$AB_2X_4$ ($A$ = Mn, Fe, Co; $B$ = Al, Sc; $X$ = O, S) spinels are frustrated magnets which exhibit spin liquid, \cite{krimmelprb_79_134406_2009spin} orbital liquid \cite{fritsch_prl_92_116401} or orbital glass \cite{fichtl_prl_94_027601_2005} states that arise from magnetic frustration effects. 
In $A$-site spinels, where the magnetic atom occupies the tetrahedrally coordinated $A$ site and forms a diamond lattice, frustration arises from competing nearest neighbour (n.n.) and next-nearest neighbour (n.n.n.) exchange interactions. \cite{tristanprb_72_174404_2005geometric,krimmelprb_79_134406_2009spin, fichtl_prl_94_027601_2005,fritsch_prl_92_116401}
This scenario is different from the case of $B$-site spinels where the magnetic atoms form a pyrochlore-type lattice and hence the observed frustration effects are inherently geometric in nature. 
Ideally, a diamond lattice antiferromagnet with only n.n. coupling $J_1$ is not magnetically frustrated. 
Additional interactions in the form of n.n.n. couplings, $J_2$, are necessary to create frustration. Monte Carlo (MC) simulations based on the "order by disorder'' \cite{villain_jp_41_1263_1980order} scenario in diamond lattice antiferromagnets predict that exotic phases like a {\it spiral spin liquid} are realized for $\left(\frac{J_2}{J_1}\right) >$ $\frac{1}{8}$.\cite{bergmannature_3_487_2007order} 
The experimental verification of {\it spiral spin liquid} in this class of materials is still lacking. 
In the Al-based $A$-site spinel series, CoAl$_2$O$_4$ has been investigated in detail as a promising candidate for the "order by disorder'' physics. 
However, existing literature on CoAl$_2$O$_4$ is contradictory -- an early report by Roth predicts an antiferromagnetic (AFM) ground state; \cite{roth_25_1_1964} Krimmel {\it et al.,} showed a liquid-like magnetic structure factor \cite{krimmelprb_79_134406_2009spin} and Tristan {\it et al.,} suggested spin-glass-like physics. \cite{tristanprb_72_174404_2005geometric} 
Recent neutron diffraction studies combined with MC simulations confirm the ground state to be AFM and the phase above $T_N$ to be a "spin liquid''. \cite{zaharkoprb_81_064416_2010evolution} 
The magnetic ground state of FeAl$_2$O$_4$ is reported to be similar to that of CoAl$_2$O$_4$ with predominantly spin-glass-like features. \cite{tristanprb_72_174404_2005geometric} 
On the other hand, MnAl$_2$O$_4$ shows long-range magnetic order below $T_N \approx$ 45~K. \cite{tristanprb_72_174404_2005geometric,krimmelprb_79_134406_2009spin} 
The existing reports on (Fe, Mn)Al$_2$O$_4$ more-or-less confirm spin-glass-like character for the Fe-spinel while the Mn-compound is described as an antiferromagnet. 
However, being in close proximity to the "spin liquid'' system CoAl$_2$O$_4$ in terms of the energy scales (comparable Curie-Weiss temperatures), it is worthwhile to re-investigate the Fe- and Mn-spinels in order to probe their physics in detail, keeping in mind that the nature of short-range magnetic correlations or the exact magnetic ground state of these systems are not unambiguously determined. 
In this paper we focus on the study of spin-spin correlations in FeAl$_2$O$_4$ and MnAl$_2$O$_4$. 
Our experiments bring us closer to the true ground state of these frustrated magnets: we show that in FeAl$_2$O$_4$ short-range magnetic order prevails  and it does not form 
a canonical spin-glass, we determine the magnetic structure of the long-range ordered phase in MnAl$_2$O$_4$ and show that short- and long-range order coexist in this 
material. 
\section{Experimental Details}
\label{exp}
The polycrystalline samples used in the present investigation were prepared through solid state synthesis routes following a prescription reported earlier. 
\cite{krimmelprb_79_134406_2009spin} 
This protocol of synthesis adopts a slow cooling rate for synthesis and thereby minimizes cation inversion. 
The cooling rates adopted for our samples were 15--10~C/h. The synthesized samples were characterized by laboratory x rays using a Huber G670 diffractometer with 
monochromatic Cu-K$\alpha$ radiation. 
\begin{figure}[!t]
\centering
\includegraphics[scale=0.42]{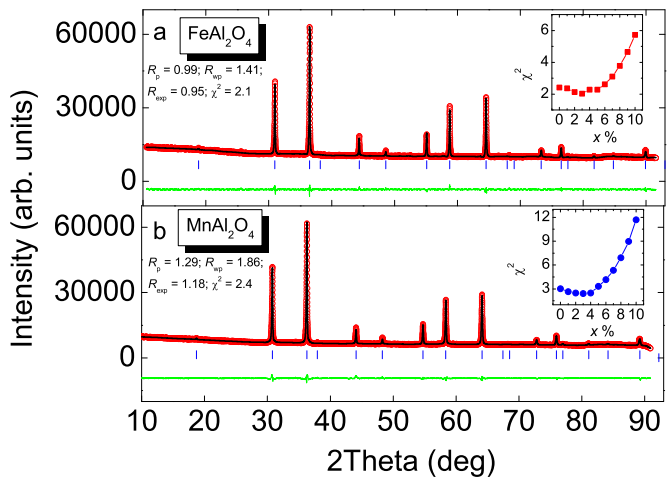}
\caption{(color online) The powder x ray diffractograms of (a) FeAl$_2$O$_4$ and (b) MnAl$_2$O$_4$ along with Rietveld refinement fits. Both the compounds are refined in cubic $Fd\overline{3}m$ space group. The reliability factors of the refinements are indicated in the graphs. The insets of (a) and (b) show the variation of goodness-of-fit with cationic inversion parameter, $x$, for FeAl$_2$O$_4$ and MnAl$_2$O$_4$ respectively.}
\label{fig_xrd}
\end{figure}
\begin{table}[!b]
\caption{The refined lattice constant ($a$), oxygen position ($u$) and inversion parameter ($x$) for FeAl$_2$O$_4$ and MnAl$_2$O$_4$ at room-temperature. The goodness-of-fit are also represented. In the cubic $Fd\overline{3}m$ space group of $AB_2X_4$, $A$-site is $8a$($\frac{1}{8}$,$\frac{1}{8}$,$\frac{1}{8}$); $B$-site is $16d$($\frac{1}{2}$,$\frac{1}{2}$,$\frac{1}{2}$) and $X$-site is $32e$($u$,$u$,$u$).}
\begin{tabular}{cccccc} \hline
Spinel                  & a (\AA)    & $u$      & $x$ & $\chi^2$ \\ \hline
FeAl$_2$O$_4$ & 8.15826(4) & 0.2644(2)  & 0.02(3)    & 2.2  \\
MnAl$_2$O$_4$ & 8.20979(4) & 0.2659(1)  & 0.014(4)  & 2.4  \\ \hline
\end{tabular}
\label{tab_str}
\end{table}
Magnetic measurements in field-cooled (FC) and zero field-cooled (ZFC) modes were carried out on polycrystalline pellets in a commercial SQUID Magnetic Property Measurement System (MPMS, Quantum Design Inc.) using ultra-low field option and also using the Vibrating Sample Magnetometer (VSM) option of Physical Property Measurement System (PPMS, Quantum Design Inc.). 
The protocol for {\it memory effect} reported by Mathieu {\it et al.,} \cite{mathieu_prb_63_092401_2001} was followed to search for a possible spin-glass phase in FeAl$_2$O$_4$. 
Neutron diffraction patterns with polarized neutrons and polarization analysis were recorded at the diffuse neutron spectrometer DNS \cite{schweika_physica_2001instrument} 
at the MLZ in Garching, Germany.  
Using XYZ polarization analysis \cite{scharpf_pss_135_359_1993xyz} the coherent nuclear, the spin-incoherent and the magnetic scattering can be separated. 
The wavelength of incident neutrons was $\lambda$ = 4.2~$\mathrm{\AA}$ leading to a maximal scattering vector magnitude of $\mathbf{Q}$ = 2.67~$\mathrm{\AA^{-1}}$. 
Due to the long neutron wavelength, only a small portion of reciprocal space can be probed. Rietveld refinements \cite{rietveld} of the x ray and neutron powder diffraction data 
were performed using FULLPROF suite program. \cite{carvajal} 
The software SARA$h$ \cite{wills_sarah} was used for magnetic structure determination using representational analysis.
\section{Results}
\subsection{Crystal Structure: FeAl$_2$O$_4$ and MnAl$_2$O$_4$}
The powder diffraction patterns of FeAl$_2$O$_4$ and MnAl$_2$O$_4$ are shown in Fig~\ref{fig_xrd} (a) and (b) respectively. Rietveld refinement of the data yield a lattice 
constant value of $a (\mathrm{\AA})$ = 8.1582(4) for FeAl$_2$O$_4$ and 8.2097(2) for MnAl$_2$O$_4$ respectively, which are comparable to reported values. 
\cite{tristanprb_72_174404_2005geometric} 
\begin{figure}[!t]
\centering
\includegraphics[scale=0.15]{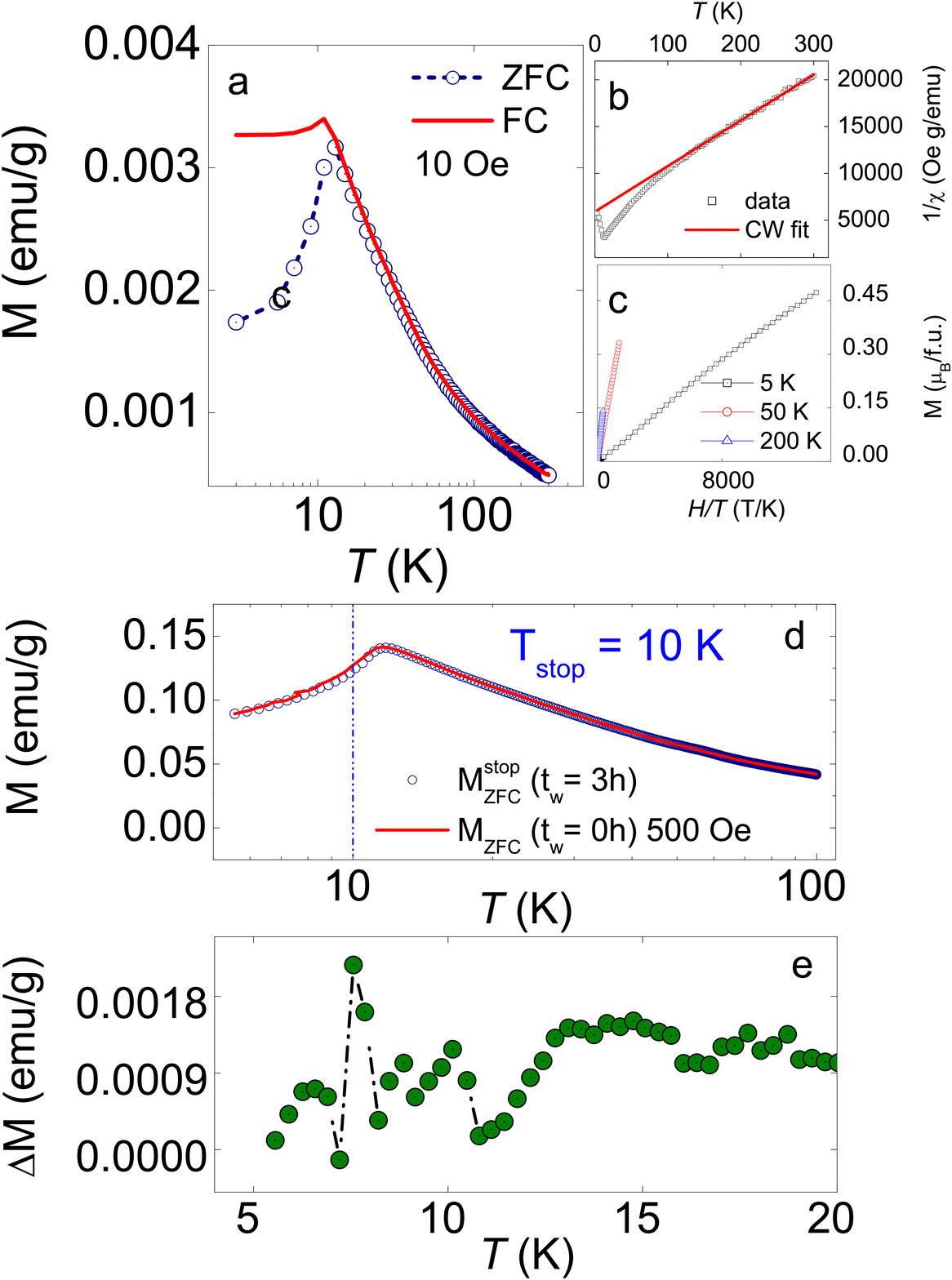}
\caption{(color online) (a) ZFC and FC magnetization curves for FeAl$_2$O$_4$ show a magnetic anomaly at $T_a \approx$13~K. (b) Shows the plot of 1/$\chi (T)$ and the CW fit which starts to deviate close to 100~K. (c) Shows that the scaling of $\mathrm{M}$ versus $\mathrm{H/T}$ deviates at low temperature ruling out the presence of superparamagnetic clusters. (d) The ZFC {\it memory effect} experiment with a $T_{stop}$ at 10~K to test the spin-glass phase in FeAl$_2$O$_4$. (e) Shows the difference between the two ZFC curves in (d) which, ideally, should display a significant peak at $T_\mathrm{stop}$ if canonical spin-glass phase was present. The error bars were comparable to the size of data points.}
\label{fig_memory_feal2o4}
\end{figure}
The results of Rietveld refinement for both spinels are shown in Fig~\ref{fig_xrd}. 
In a "normal'' $AB_2X_4$ spinel with cubic $Fd\overline{3}m$ (227) space group, $A$ occupies the Wyckoff position $8a$ ($\frac{1}{8}$, $\frac{1}{8}$, $\frac{1}{8}$);  $B$ 
occupies $16d$ ($\frac{1}{2}$, $\frac{1}{2}$, $\frac{1}{2}$) and oxygen occupies $32e$ ($u$, $u$, $u$) where $u$ = 0.25. 
The refined value of $u$ = 0.2644(2) obtained in the present case for FeAl$_2$O$_4$ deviates from the ideal spinel value but is comparable with previous reports claiming a low 
degree of inversion. \cite{tristanprb_72_174404_2005geometric} 
In order to ascertain the percentage of cation inversion, refinement of the x ray data was carried out assuming the stoichiometric formula 
($A_{1-x}$Al$_{x}$)[Al$_{2-x}A_{x}$]O$_4$ ($A$ = Fe, Mn) where $x$ is the inversion parameter. 
During the Rietveld fit of the x ray data, the parameters that were varied are the zero point shift, scale factor, background, profile parameters, lattice parameters, oxygen 
position and the inversion parameter. 
The Debye-Waller factors were fixed at isotropic values obtained from powder diffraction studies on similar spinels.\cite{oneill_pcm_18_302_1991}
A value of the order of $x$ = 0.02(3) was obtained suggesting that the degree of inversion is insignificant. 
The Rietveld refinements were performed to test for the best value of $\chi^2$ by varying the degree of inversion from $x$ = 0 to 1.
The results obtained in this way support the low content of inversion in the samples and are presented as insets of fig~\ref{fig_xrd} (a) and (b) for FeAl$_2$O$_4$ and 
MnAl$_2$O$_4$ respectively.
Within the resolution of the observed data, no residues of unreacted oxides of Fe or Mn were traceable however an unidentified weak signal at 2$\Theta \sim$ 26$^{\circ}$ is 
observed in the case of FeAl$_2$O$_4$. 
The magnetic measurements which are presented later on do not lead to suspicion about major magnetic impurities in the samples. 
In Figure~\ref{fig_memory_feal2o4} (a), which shows the ZFC and FC magnetization curves for FeAl$_2$O$_4$, a magnetic anomaly is evident at 13~K, where the curves 
bifurcate. 
The magnetic susceptibility at high temperature was analysed using the Curie-Weiss expression, $\chi (T) = \frac{\mu^2_B}{3k_B} \left( \frac{g^2_\mathrm{eff}S(S + 1)}{T - 
\Theta}\right)$ leading to  an effective paramagnetic moment, $\mu_\mathrm{eff}$ = $g_\mathrm{eff}\sqrt{S(S + 1)}\approx$ 5.3(4)~$\mu_\mathrm{B}$/f.u. and 
Curie-Weiss temperature, $\Theta_\mathrm{CW}$ = -120(1)~K. 
%
%
The effective paramagnetic moment is comparable to the spin-only moment value 4.9~$\mu_\mathrm{B}$/f.u. for Fe$^{2+}$ (3$d^6$; $S$ = 2; $L$ = 2).
Slight differences in the effective moment value in the case of FeAl$_2$O$_4$ have been attributed to the orbital degrees of freedom pertaining to 
Fe$^{2+}$.\cite{tristanprb_72_174404_2005geometric,krimmelprb_79_134406_2009spin}
\\
Previous reports of bulk magnetic properties, neutron diffraction and M\"{o}ssbauer spectroscopy have attributed the observed anomalies to a spin-glass or spin-glass-like phase in 
FeAl$_2$O$_4$. \cite{tristanprb_72_174404_2005geometric,soubeyroux_ljpc_49_C8_1988spin} 
\begin{figure}[!b]
\centering
\includegraphics[scale=0.25]{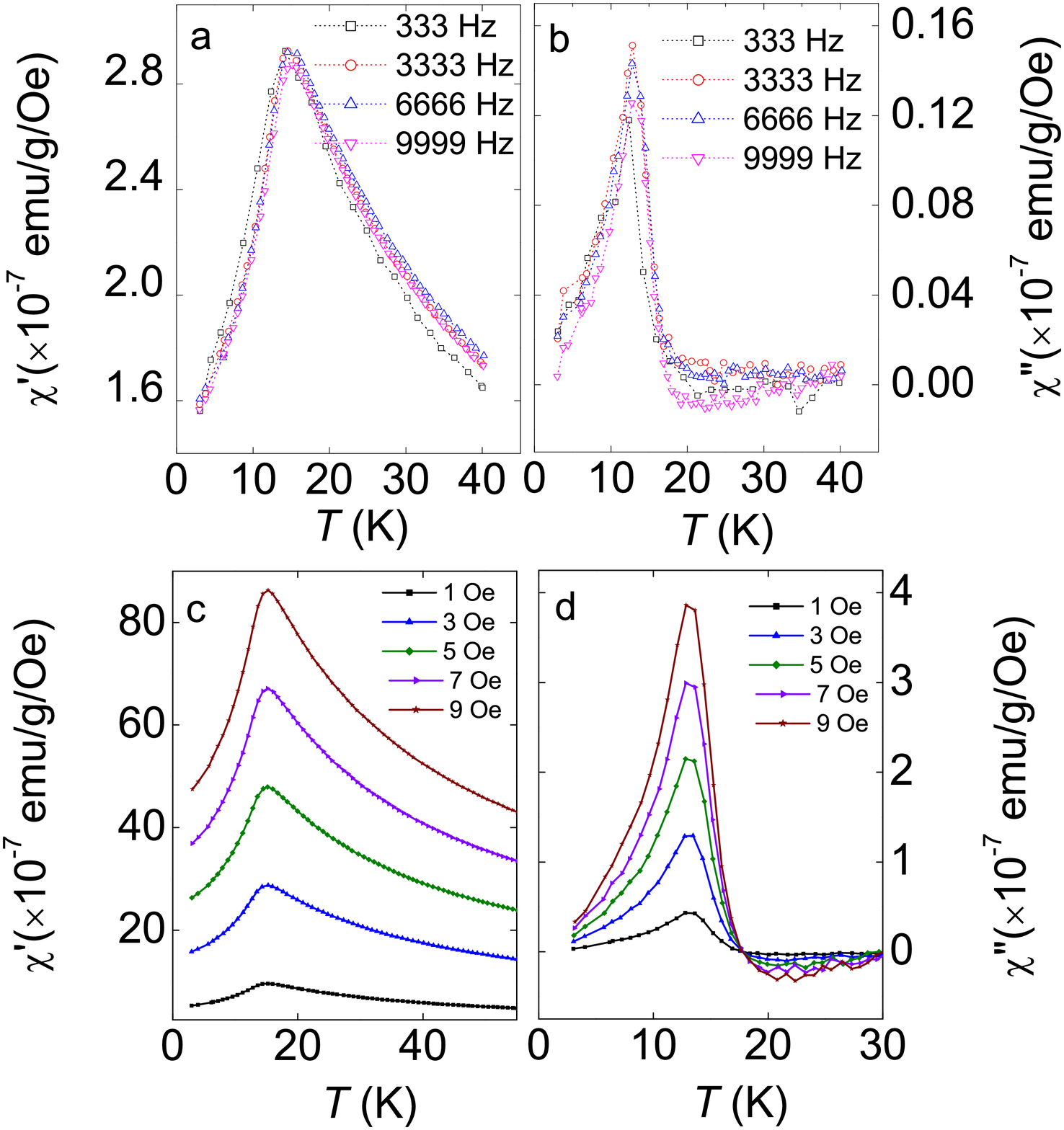}
\caption{(color online) Real (a) and imaginary (b) parts of ac susceptibility, $\chi (T)$, of FeAl$_2$O$_4$ at various frequencies. No significant frequency-dependence of $T_a$ is observed. Real (c) and imaginary (d) parts of $\chi (T)$ at 9999~Hz with increasing dc amplitude of measuring ac field. For the sake of clarity, only the curves at 2~Oe interval are shown.}
\label{fig_ac_feal2o4}
\end{figure}
The plot of inverse magnetic susceptibility and CW fit are presented in Fig~\ref{fig_memory_feal2o4} (b) which shows deviation from linear behaviour close to 100~K itself, 
signifying that the spin correlations extend to high temperatures. 
In Fig~\ref{fig_memory_feal2o4} (c), a scaling of $\mathrm{M}$ versus $\mathrm{H/T}$ at three different temperatures, derived from the isothermal magnetization plots, is 
presented. 
At low temperatures below the $T_a$, a perfect scaling is not observed thus ruling out the possibility that FeAl$_2$O$_4$ could be a system of non-interacting paramagnetic 
clusters.\cite{bean_jap_27_1448_1956magnetic} 
\\
In order to ascertain whether a genuine canonical spin-glass phase is present, we performed {\it memory effect} experiments. 
In this experiment, initially a ZFC curve $M_\mathrm{zfc}$ was recorded in the warming cycle at constant rate of heating. 
Next, the sample was zero field-cooled to $T_\mathrm{stop} = 10~K < T_a$ and held constant for $t_w$ = 3~h; afterwards the sample is cooled to the lowest temperature. 
Finally, the magnetization $M^\mathrm{stop}_\mathrm{zfc}$ was measured while warming the sample at the same constant rate as done for $M_\mathrm{zfc}$. 
The resulting magnetization curves are presented in Fig~\ref{fig_memory_feal2o4} (d). For canonical spin-glass materials, the difference between these two curves, {\it i.e.,} $\Delta M$ = ($M_\mathrm{zfc}$ - $M^\mathrm{stop}_\mathrm{zfc}$) should present a peak centered at the temperature at which the halt was administered ({\it i.e.,} $T_\mathrm{stop}$). \cite{mathieu_prb_63_092401_2001} 
The $\Delta M$ for FeAl$_2$O$_4$ presented in Fig~\ref{fig_memory_feal2o4} (e) does not support a canonical spin-glass state. 
The {\it memory effect} experiment was repeated for lower values of applied fields of 50 and 100~Oe which also did not present "memory". 
The {\it memory effect} in canonical spin-glasses can be described based on heirarchical organization of metastable states as a function of temperature which attempts to explain the effect based on a phase space picture where the free energy barriers grow as temperature is reduced.\cite{refregier_jp_48_1533_1987ageing} 
Equivalently, a real space picture of "droplets" of spins can also be used. \cite{fisher_prb_38_386_1988equilibrium} 
Below the spin-glass transition, the "droplet-picture" assumes the presence of only one phase and it's spin-reversed counterpart but the spins arrange randomly as dictated by the disordered nature of interactions and effect of temperature. 
Though the spinel compounds investigated in the present work are frustrated, the lack of site disorder reduces the probability of formation of a true spin-glass phase.
\\
Ac susceptibility experiments on FeAl$_2$O$_4$ also do not support a spin-glass-like phase. 
The real and imaginary parts of ac susceptibility at different frequencies in the range 333~Hz to 9999~Hz presented in Fig~\ref{fig_ac_feal2o4} (a) and (b) do not show frequency-dependent shifts of $T_a$ which is a typical feature of spin-glasses. 
Crucial features of spin-glass state are the presence of relaxation processes of all time scales and aging. 
Signature of aging is visible in the real and imaginary parts of ac susceptibility where the susceptibility relaxes down. 
Moreover, for a canonical spin glass transition, dynamical scaling holds near the critical temperature and hence susceptibility should obey critical scaling. 
Then, the apparent spin-glass transition temperatures at finite frequencies as fixed by the peak of ac susceptibility curves must obey power law. \cite{mydosh_book} 
These features are absent in the ac susceptibility of FeAl$_2$O$_4$. 
Increasing the dc amplitude of the ac measuring field only enhances the magnetic susceptibility, typical of antiferromagnets, Fig~\ref{fig_ac_feal2o4} (c) and (d). 
\subsection{Polarized Neutron Scattering: FeAl$_2$O$_4$}
\subsubsection{Short-range spin correlations}
From the macroscopic magnetic measurements, it is clear that FeAl$_2$O$_4$ has a complex ground state which is not explainable within the canonical spin-glass scenario. 
In order to understand the observed magnetic properties and to disentangle the long- and short-range magnetic contributions, spin-spin correlation functions are extremely helpful. 
\begin{figure}[!t]
\centering
\includegraphics[scale=0.30]{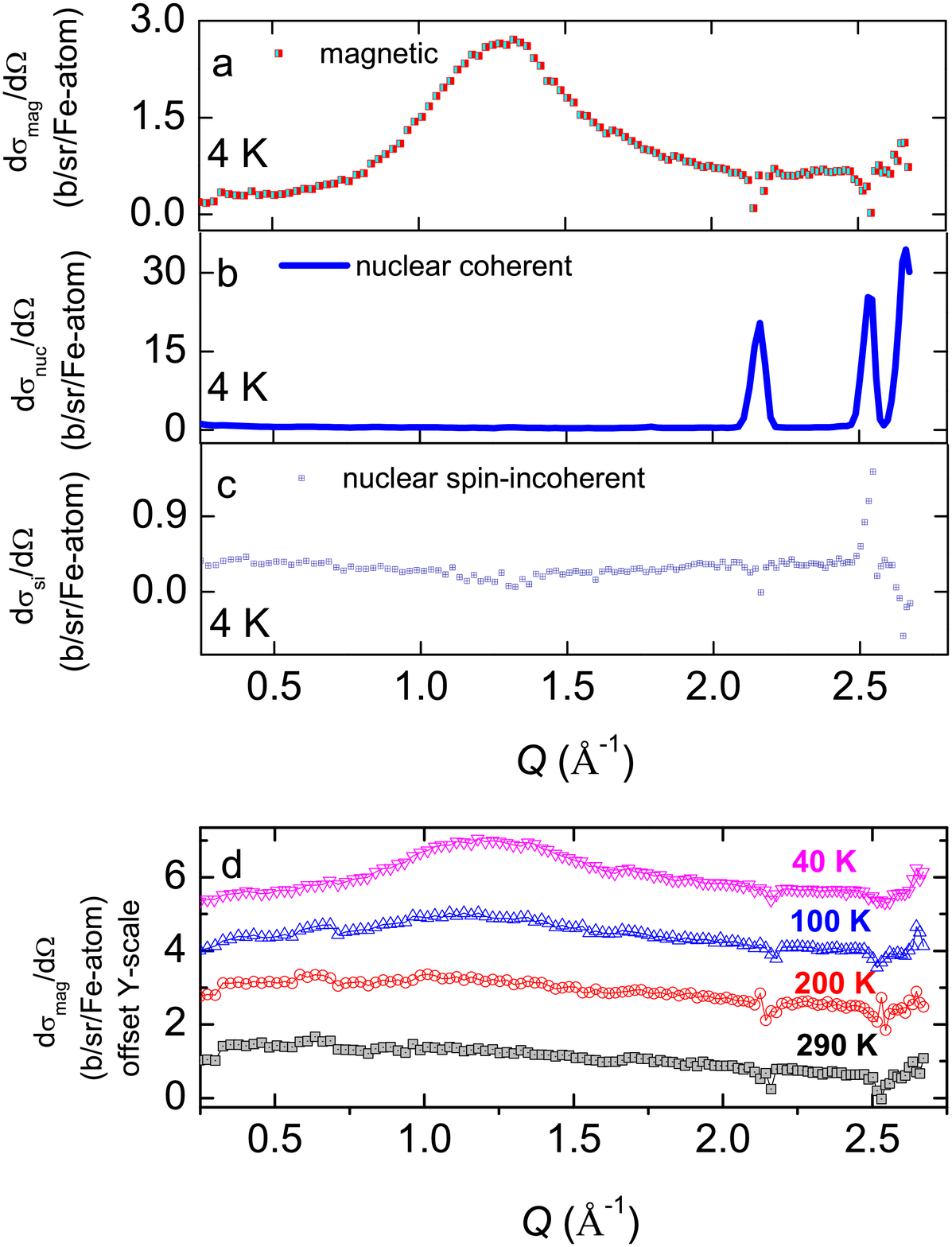}
\caption{(color online) Magnetic (a), nuclear coherent (b), and nuclear spin-incoherent (c), contributions to the differential scattering cross-section of FeAl$_2$O$_4$ at 4~K. (d) Shows the magnetic scattering cross-sections at 40, 100, 200 and 290~K. The Y-axis scale is offset by a factor of 1.2 for clarity. Polarization analysis becomes less reliable at the positions of strong nuclear reflections, which produces the scatter of data points.} 
\label{fig_dns_feal2o4}
\end{figure}
The spin correlation functions can be experimentally determined through the technique of polarized neutron scattering with polarization
analysis which separates the nuclear and magnetic scattered intensities from the total scattered intensity. 
Within the quasistatic approximation, the nuclear coherent ($nc$), nuclear spin-incoherent ($nsi$) and magnetic ($m$) scattering cross sections are separated using the $xyz$-polarization analysis in the {\it spin-flip} (SF) and {\it non spin-flip} (NSF) channels. \cite{scharpf_pss_135_359_1993xyz} 
The $xy$-plane is defined by the multi-detector array of DNS instrument which makes the scattering vector $\mathbf{Q}$ always perpendicular to the $z$-direction. 
The SF and NSF partial differential cross sections are then, ($\frac{d\sigma^{x,y,z}_{SF}}{d\Omega}$) and ($\frac{d\sigma^{x,y,z}_{NSF}}{d\Omega}$). 
By making suitable combinations of the expressions for cross sections, the nuclear coherent, nuclear spin-incoherent and magnetic scattering components can be separated \cite{scharpf_pss_135_359_1993xyz}:
\begin{align*}
\left(\frac{d\sigma}{d\Omega}\right)_{m.} & = 2\left( \frac{d\sigma^{SF}_x}{d\Omega} + \frac{d\sigma^{SF}_y}{d\Omega} - 2\frac{d\sigma^{SF}_z}{d\Omega} \right) \\
&=  -2\left( \frac{d\sigma^{NSF}_x}{d\Omega} + \frac{d\sigma^{NSF}_y}{d\Omega} - 2\frac{d\sigma^{NSF}_z}{d\Omega}\right) \\
\left( \frac{d\sigma}{d\Omega} \right)_{nc.} &= \left( \frac{d\sigma^{NSF}_z}{d\Omega} - \frac{1}{2} \frac{d\sigma}{d\Omega}_{m.} - \frac{1}{3}\frac{d\sigma}{d\Omega}_{nsi.} \right) \\
\left( \frac{d\sigma}{d\Omega}\right)_{nsi.} &= \frac{3}{2}\left( 3\frac{d\sigma^{SF}_z}{d\Omega} -  \frac{d\sigma^{SF}_x}{d\Omega} - \frac{d\sigma^{SF}_y}{d\Omega} \right) \\
\end{align*}
These equations hold for ideal experimental conditions for polarization and flipping ratio. 

In order to obtain the absolute scattering cross sections, vanadium standard is usually measured separately in the same experimental conditions. 
\\
The results of polarized neutron scattering experiment on FeAl$_2$O$_4$ are presented in Fig~\ref{fig_dns_feal2o4}. 
The magnetic (a), nuclear coherent (b) and nuclear spin-incoherent (c) contributions to differential scattering cross-section at 4~K are shown separately. 
Figure~\ref{fig_dns_feal2o4} (d) shows the magnetic cross-sections at 40, 100, 200 and 290~K where a gradual loss of spin correlations and transition to a pure form factor square decay of the intensity with temperature is evident. 
\begin{table*}
\caption{Parameters estimated from fits to magnetic cross-section of FeAl$_2$O$_4$ using Eqn~(\ref{eq_peakfit}). $g$ is the Land\'{e} factor, $\frac{\langle S_0 \cdot S_n \rangle}{S(S + 1)}$ denotes the spin-spin correlations and $\epsilon^2$ is an indicator of the goodness of fit, defined such that for the best fit $\epsilon^2 \approx$ 1. The parameters ($c_n$, $R_n (\AA)$) used for the fits, for example at 4~K, are: ($c_1, R_1$) = (4, 3.53); ($c_2, R_2$) = (12, 5.77); ($c_3, R_3$) = (12, 6.76); ($c_4, R_4$) = (6, 8.15); ($c_5, R_5$) = (12, 8.89); ($c_6, R_6$) = (8, 9.9). The terms for the third and higher shells are rather insignificant however, they were necessary for a faithful fit.}
\begin{tabular}{cccccccccc} \hline
T (K) &  $g$ & $\frac{\langle S_0 \cdot S_1 \rangle}{S(S + 1)}$ & $\frac{\langle S_0 \cdot S_2 \rangle}{S(S + 1)}$ & $\frac{\langle S_0 \cdot S_3 \rangle}{S(S + 1)}$ & $\frac{\langle S_0 \cdot S_4 \rangle}{S(S + 1)}$ & $\frac{\langle S_0 \cdot S_5 \rangle}{S(S + 1)}$ & $\frac{\langle S_0 \cdot S_6 \rangle}{S(S + 1)}$ & $\epsilon^2$  \\ \hline
4  & 2.453(18) & -0.512(20) & 0.144(15) & 0.013(20) & -0.056(71) & -0.095(37) & 0.030(14) & 0.985 \\
40 & 2.435(18) & -0.325(12) & 0.056(16) & 0.005(18) & 0.000(80) & -0.032(16) & - & 0.965 \\
100 & 2.359(17) & -0.157(12) & 0.005(6) & - & - & - & - & 0.933 \\ \hline
\end{tabular}
\label{tab1}
\end{table*}
The nuclear coherent scattering intensities of FeAl$_2$O$_4$  at 4~K as presented in Fig~\ref{fig_dns_feal2o4} (b), can consistantly be described with the structural parameters obtained from x ray refinements. The refined lattice parameters are comparable to the values obtained from x ray data analysis.
It is clear from Fig~\ref{fig_dns_feal2o4} (a) that FeAl$_2$O$_4$ does not enter a magnetically long-range ordered state down to 4~K. 
The broad and diffuse nature of the magnetic scattered intensity suggests that the spins in FeAl$_2$O$_4$ are correlated to each other over short distances. 
\\
In order to obtain a quantitative estimate of the spin-spin correlations in FeAl$_2$O$_4$, we modeled the observed spin correlations by calculating the powder averaged differential magnetic cross-section, \cite{rainford_jphy_43_c7_1982} 
\begin{eqnarray}
\left(\frac{d\sigma_{mag}}{d\Omega}\right) = \frac{2}{3} \left(\frac{\gamma e^2}{mc^2}\right)^2 \left(\frac{g}{2}\right)^2 S(S+1) f^2(\mathbf{Q}) \times \nonumber \\
\left[ 1 + \sum^N_{n=1} \frac{\langle S_0 \cdot S_n \rangle}{S(S + 1)}c_n \frac{\sin (QR_n)}{QR_n} \right]
\label{eq_peakfit}
\end{eqnarray}
where, ($\gamma e^2/m c^2$) = 5.39~pm is the magnetic scattering length, $f (\mathbf{Q})$ is the magnetic form factor of the Fe$^{2+}$ ion (or Mn$^{2+}$ in the case of MnAl$_2$O$_4$) \cite{brown_1995magnetic} and $R_n$ and $c_n$ are the radius and coordination number of the $n$th near-neighbour shell respectively. 
The above expression is a simplified form for isotropic spin pair correlations, while the complete form contains additional terms, see [\onlinecite{blech_physics_1_31_1964}]. 
\begin{figure}[!t]
\centering
\includegraphics[scale=0.35]{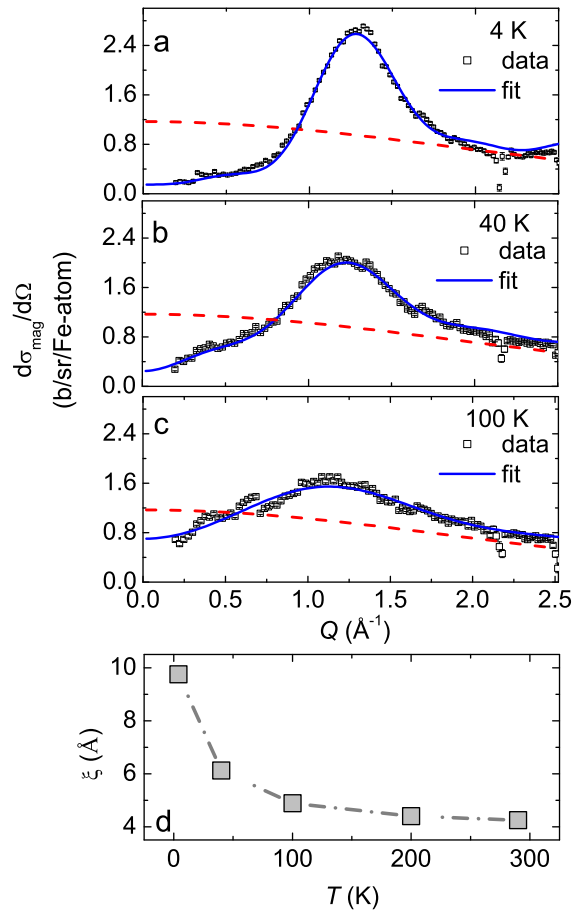}
\caption{(color online) (a-c) The magnetic scattering intensity of FeAl$_2$O$_4$ at 4, 40 and 100~K along with the fit curve (blue solid line) assuming 4$^\mathrm{th}$ n.n. (40~K) shell and 5$^{th}$ n.n. (4~K) shell, following Eq~(\ref{eq_peakfit}). The red ($--$) line is drawn proportional to $f^2(\mathbf{Q})$ for Fe$^{2+}$. (d) shows that the spin-spin correlation length, $\xi$, decreases as temperature increases.}
\label{fig_peakfit_feal2o4}
\end{figure}
The magnetic part of the differential cross-section data for FeAl$_2$O$_4$ at 4, 40 and 100~K were least-square fitted as per Eqn~(\ref{eq_peakfit}). 
The normalized $n^\mathrm{th}$ near-neighbour pair correlations, $\frac{\langle S_0 \cdot S_n \rangle}{S(S + 1)}$, and the Land\'{e} factor, $g$, were set as adjustable fitting parameters. 
The initial value of $S$ was fixed at 2 corresponding to the spin of Fe$^{2+}$. 
The fitting results are shown as solid lines in Fig~\ref{fig_peakfit_feal2o4} (a), (b) and (c) along with the data for 4, 40 and 100~K. 
The fit parameters are summarized in Table~\ref{tab1}. For a comparison, the form-factor of Fe$^{2+}$ ($S$ = 2) is plotted in Fig~\ref{fig_peakfit_feal2o4}.
The fitting is limited to $n$ = 6, 5 and 2 for temperatures 4, 40 and 100~K, respectively, as can be seen from the table. 
The negative and positive signs of the normalized near-neighbour pair correlations stand for antiferromagnetic and ferromagnetic type of spin correlations, respectively. 
It is clear that the nearest neighbours are antiferromagnetically coupled and the next-nearest-neighbours show weak ferromagnetic correlations. 
There exist very little correlations for $n >$ 3 neighbours, but including the $n >$ 3 neighbours was necessary to achieve good fits to the 4 and 40~K data. 
Also, we can see the degree of correlation decreases as the temperature increases from 4 to 100~K.
 The Land\'{e} factor, $g$, estimated from the fit is larger than the spin-only value 2, which could be due to orbital contribution from Fe$^{2+}$ ions. 
 This result may suggest an orbital contribution to the estimated effective moment as deduced from the CW analysis of the magnetic susceptibility. 
 The magnetic scattering intensity at different temperatures, as presented in Fig~\ref{fig_dns_feal2o4}, were also analyzed using Lorentzian function to model the peak function and extract the magnetic correlation length, $\xi$.  
 The value, $\xi (\mathrm{\AA}) \approx$ 11 obtained is roughly three times the Fe--Fe bond distance in FeAl$_2$O$_4$ ($\approx$ 3.5~$\mathrm{\AA}$), or, close to fifth n.n distances of Fe (A preliminary estimate using the expression $\xi \approx$ $\left( \frac{2\pi}{\Delta \mathbf{Q}} \right)$, where $\Delta \mathbf{Q}$ is the full-width-at-half-maximum of the correlation peak also gave similar value). 
 In Fig~\ref{fig_peakfit_feal2o4} (d), the temperature evolution of $\xi$ is presented, which shows that the correlation length gradually decreases with temperature. 
 It is clear that the short-range spin-spin correlations persist to temperatures $T > T_a$.
\subsection{Magnetization: MnAl$_2$O$_4$}
\begin{figure}[!b]
\centering
\includegraphics[scale=0.26]{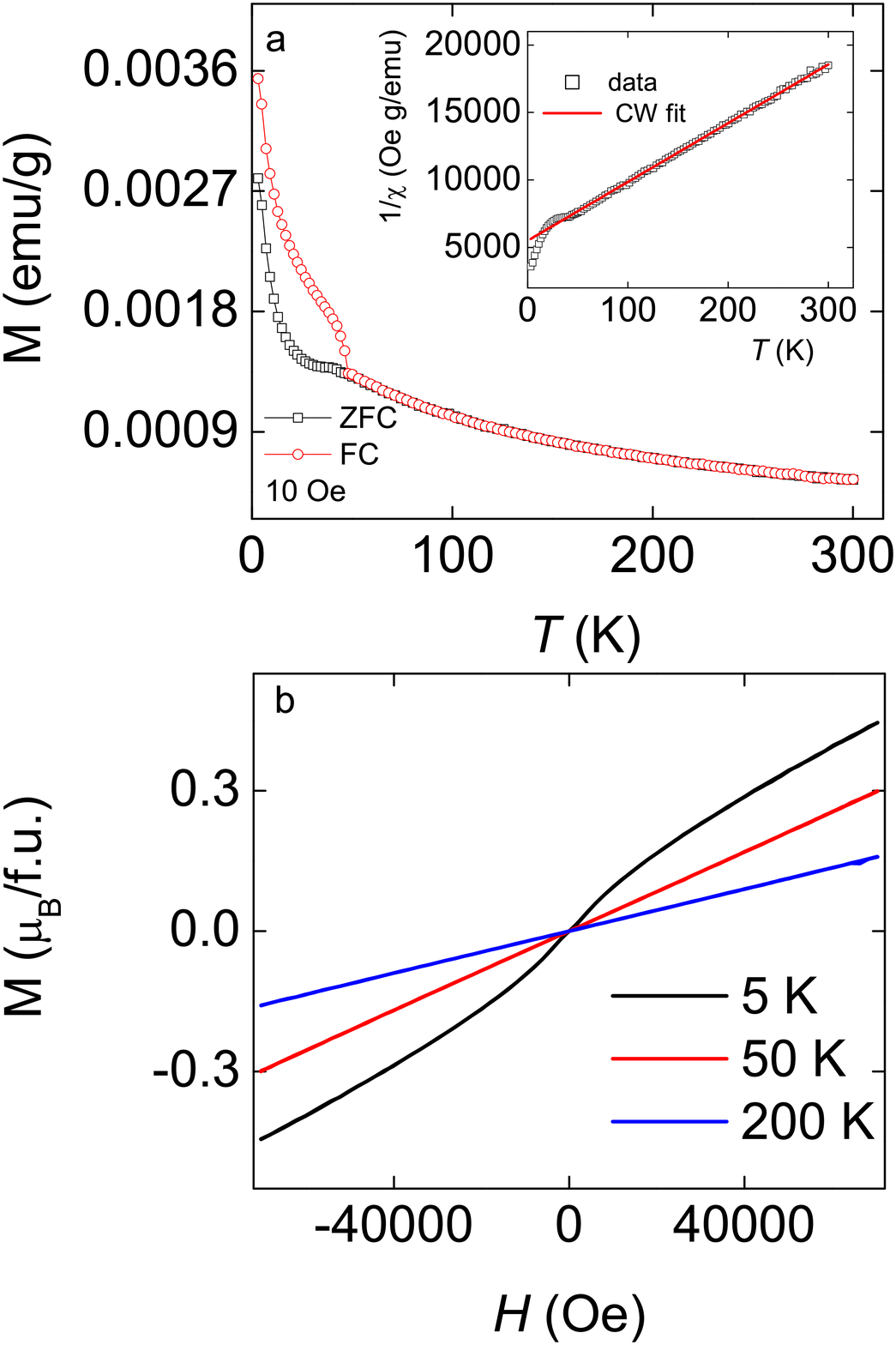}
\caption{(color online) (a) The ZFC and FC magnetization curves of MnAl$_2$O$_4$ showing a bifurcation at $T_N \approx$ 40~K. The inset shows the Curie-Weiss fit of inverse magnetic susceptibility. (b) Shows the magnetic hysteresis at 5, 50 and 200~K.}
\label{fig_mt_mnal2o4}
\end{figure}

The magnetization data of MnAl$_2$O$_4$ presented in Fig~\ref{fig_mt_mnal2o4} display a magnetic phase transition at $\approx$ 40~K. 
Analyzing the paramagnetic regime of the magnetization curve using the CW law, an effective paramagnetic moment $\mu_{\mathrm{eff}}\sim$ 
5.6(3)~$\mu_\mathrm{B}$/f.u. and $\Theta_\mathrm{CW}$ = -126(2)~K are obtained. 
The experimentally obtained effective moment value is comparable to the theoretical spin-only value of Mn$^{2+}$ (3$d^5$; $S$ = $\frac{5}{2}$; $L$ = 0), 
5.9~$\mu_\mathrm{B}$/f.u. 
The frustration parameter $f$ for MnAl$_2$O$_4$ is 3.5, which is low compared to the value for FeAl$_2$O$_4$. Though the $f$ value is less than half of that for FeAl$_2$O$_4$, 
the $\Theta_\mathrm{CW}$ values of both systems are comparable, indicating that the dominant exchange interactions are comparable.
\subsection{Polarized Neutron Scattering: MnAl$_2$O$_4$}
\subsubsection{Co-existence of short- and long-range spin correlations}
\begin{figure}[!b]
\centering
\includegraphics[scale=0.4]{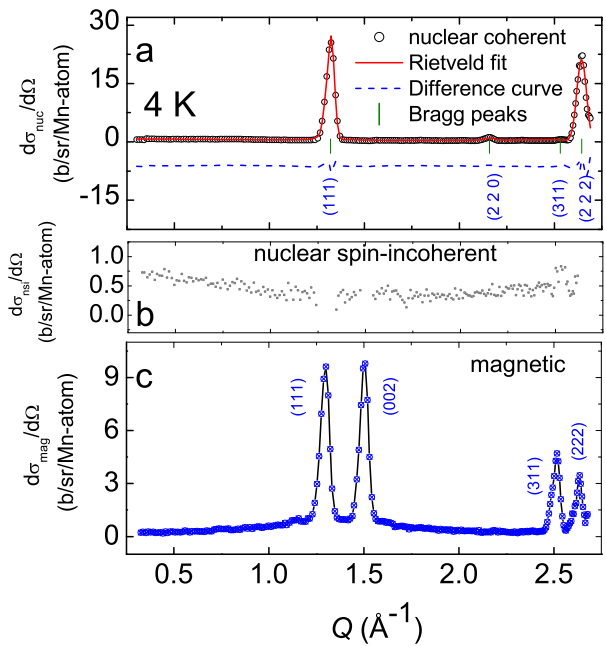}
\caption{(color online) (a) Nuclear coherent contribution to the differential scattering cross-section of MnAl$_2$O$_4$ at 4~K. The results of Rietveld fit in $Fd\overline{3}m$ space group are shown and Bragg peaks are marked. (b) Shows  the nuclear spin-incoherent cross section.  (c) Shows the indexed magnetic peaks at 4~K. In contrast to FeAl$_2$O$_4$, long-range order is present at 4~K.}
\label{fig_dns_mnal2o4_a}
\end{figure}

\begin{figure}[!t]
\centering
\includegraphics[scale=0.4]{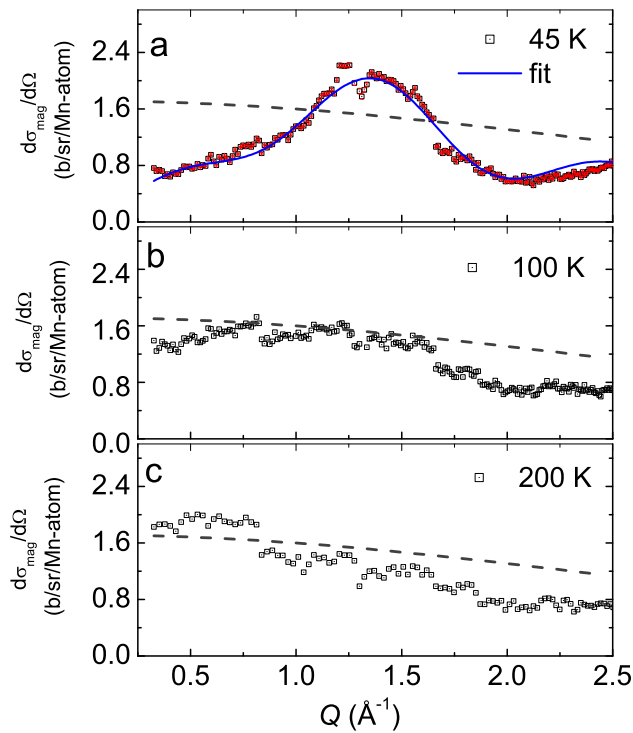}
\caption{(color online) (a-c) Magnetic scattering cross-sections at 45, 100 and 200~K, respectively. The curve-fit according to Eqn~(\ref{eq_peakfit}) for 45~K is shown. The step-like features seen in 200~K data are attributable to inefficient detector calibration. The gray ($--$) line is drawn proportional to $f^2(\mathbf{Q})$ of Mn$^{2+}$.}
\label{fig_dns_mnal2o4_b}
\end{figure}

\begin{figure}[!b]
\centering
\includegraphics[scale=0.4]{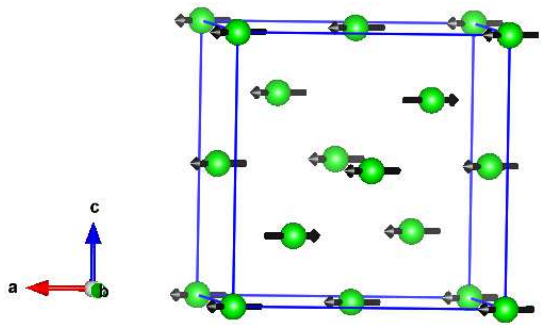}
\caption{(color online) The magnetic structure of MnAl$_2$O$_4$. Only the Mn atoms that form the diamond lattice are shown. The figure was created using the software {\it Balls $\&$ Sticks}.\cite{ballsandsticks} Note that the moment direction cannot be determined from neutron powder diffraction for a cubic crystal system.}
\label{fig_structure}
\end{figure}

The results of polarized neutron scattering experiment on MnAl$_2$O$_4$ are presented in Fig~\ref{fig_dns_mnal2o4_a} and Fig~\ref{fig_dns_mnal2o4_b}. 
In Fig~\ref{fig_dns_mnal2o4_a} (a), the nuclear coherent contribution to the differential scattering cross-section at 4~K is shown. The structural model as obtained from x ray powder diffraction yields reasonable agreement, while the DNS instrument is not a dedicated powder diffractometer. 
Similar to the case of FeAl$_2$O$_4$, the nuclear scattering intensity was refined using the $Fd\overline{3}m$ space group symmetry, as presented in Fig~\ref{fig_dns_mnal2o4_a} (a) for 4~K data. The nuclear spin-incoherent and magnetic contributions at 4~K are presented in Fig~\ref{fig_dns_mnal2o4_a} (b, c) respectively. 
The separated magnetic intensity exhibits (111) and (222) Bragg peaks and additional peaks at (002) and (311). The magnetic reflections could be indexed using $\bf{k}$ = (000) which lead to two possible irreducible representations (IR), $\Gamma_8$ and $\Gamma_9$ (according to the numbering scheme of Kovalev \cite{kovalev1993}). Though limited in $\mathbf{Q}$-range, the magnetic scattered intensity was analyzed in $\Gamma_8$ representation and the results of the Rietveld refinement are presented in Fig~\ref{fig_dns_mnal2o4_a} (c) as the indices of magnetic reflections $(1 1 1)$, $(0 0 2)$, $(3 1 1)$ and $(2 2 2)$. 
The magnetic structure suggested by the refinement is schematically represented in Fig~\ref{fig_structure}. 
However, the magnetic moment direction in the case of cubic symmetry cannot be uniquely determined from powder diffraction data alone. \cite{shirane_acta_12_282_1959note} 
The observed and calculated intensities for  $(1 1 1)$, $(0 0 2)$, $(3 1 1)$ and $(2 2 2)$ reflections are collected in Table~\ref{tab3} for a comparison. 
The average magnetic moment at 4~K visible in the Bragg reflection is refined to a value of 1.17(4)~$\mu_\mathrm{B}$ which is considerably reduced compared to the spin-only saturation moment assuming complete long range order of the Mn-lattice.
Roth reported the magnetic structure of MnAl$_2$O$_4$ where each spin at the $A$ site is surrounded by oppositely aligned nearest neighbour spins in one tetrahedra.
\cite{roth_25_1_1964} 
However, Tristan {\it et al.,} \cite{tristanprb_72_174404_2005geometric} point out that the simple collinear picture and the estimated transition temperature of 6.4~K by Roth disagree with the observation of diminished value of local moment 5~$\mu_{\mathrm{B}}$ (for a fully ordered spin-only system)
and $T_N \approx$ 40~K. 
Thus it is clear that significant magnetic disorder persist down to lowest temperatures. 
The cation inversion was suggested as the reason where Mn spins at $A$ and $B$ sites build paramagnetic clusters. 
However, later, by neutron powder diffraction study the magnetic structure of MnAl$_2$O$_4$ was deduced as a collinear antiferromagnetic type with $T_N \approx$ 42~K.\cite{krimmelphysica_378_583_2006spin}
\\
It is clear that the low temperature magnetic properties of MnAl$_2$O$_4$ are different from those of FeAl$_2$O$_4$ as sharp peaks signify long-range magnetic order to emerge below 45~K (Fig~\ref{fig_dns_mnal2o4_a} (c)) . 
However, besides sharp peaks due to long-range order, we observe diffuse intensity in the ordered state, which survives above the Ne\'{e}l temperature $T_N \approx$ 40~K, centered around $|\mathbf{Q}|\approx$ 1.3~$\mathrm{\AA^{-1}}$. 
At 200~K, the scattered intensity follows more or less the form factor behaviour of free magnetic ion Mn$^{2+}$. The magnetic scattering cross-section of MnAl$_2$O$_4$ at 45, 100 and 200~K is presented in Fig~\ref{fig_dns_mnal2o4_b} (a-c). 
The diffuse magnetic intensity observed for MnAl$_2$O$_4$ at 45~K and 100~K were analyzed using Eqn~(\ref{eq_peakfit}) as was done in the case of FeAl$_2$O$_4$. 
The results of the fit are presented in Fig~\ref{fig_dns_mnal2o4_b} (a) as solid lines and the estimated parameters for 45~K data are presented in Table~\ref{tab3}. 
Inclusion of three n.n shells was required to obtain a satisfactory fit. 
Since Mn$^{2+}$ is an $L$ = 0 ion, the deviation from the spin-only value is not significant.
Fits were also performed by keeping $g$ constant at the spin-only value, however, it did not lead to significant changes in the parameters.
 From the refined parameters, $\frac{\langle S_0 \cdot S_1 \rangle}{S(S + 1)}$, prominant n.n antiferromagnetic and n.n.n ferromagnetic exchange were concluded, similar to the case of FeAl$_2$O$_4$. 
 In the case of FeAl$_2$O$_4$, $\frac{\langle S_0 \cdot S_1 \rangle}{S(S + 1)}$ = -0.325(12) at 40~K whereas in the case of MnAl$_2$O$_4$, it is -0.360(10) at 45~K. 
\\
The temperature evolution of the peak intensity of the $(002)$ magnetic reflection of MnAl$_2$O$_4$ is presented in Fig~\ref{fig_magT_both} (a). 
For this measurement, the detector bank of the DNS instrument was aligned such that $(002)$ Bragg peak illuminates a specific detector tube and the intensity was recorded as a function of temperature. 
\begin{table}[!b]
\caption{Spin-spin correlation parameters estimated from the fits to the magnetic cross-section of MnAl$_2$O$_4$ using Eqn~(\ref{eq_peakfit}) for 45~K data. }
\begin{tabular}{cccccccccc} \hline\hline
$g$ & $\frac{\langle S_0 \cdot S_1 \rangle}{S(S + 1)}$ & $\frac{\langle S_0 \cdot S_2 \rangle}{S(S + 1)}$ & $\frac{\langle S_0 \cdot S_3 \rangle}{S(S + 1)}$ & $\epsilon^2$ \\ \hline
2.110(13) & -0.360(10) & 0.166(10) & -0.122(11) & 0.953 \\ \hline\hline
\end{tabular}
\label{tab2}
\end{table}
As such, the detector signal does not represent the integrated but the peak intensity. 
The angular coverage of a single tube is sufficient to account for the lattice expansion in the respective temperature region. 
The magnetic intensity sharply decreases around $T_N \approx$ 40~K where the magnetic phase transition occurs, however, it can be seen from the figure that even at $T > T_N$, there exists non-zero intensity signifying the presence of magnetic fluctuations. 
We analyse the temperature evolution in terms of a mean-field approximation, giving the reduced sublattice magnetization \cite{smart1966}: 
\begin{equation}
M=B_S \left[\frac{3S}{(S+1)} \left(\frac{T_N}{T}\right)M \right]
\label{eqn_Bfit}
\end{equation}
where $M$ is the magnetic intensity, $S$ is the spin quantum number and $B_S$ is the Brillouin function. 
The scattered intensity is proportional to the square of the magnetization. 
We take the diffuse intensity into account by adding a sloping line. Leaving $S$ a free parameter, the model yields $S = 2.3(2), T_N = 40(1)$ K in agreement with $S = \frac{5}{2}$ for Mn$^{2+}$ and the macroscopic data.
\section{Discussion}
Previous reports on magnetic properties of $A$-site spinels have suggested spin-glass-like and antiferromagnetic ground states, respectively, for FeAl$_2$O$_4$
and MnAl$_2$O$_4$. \cite{tristanprb_72_174404_2005geometric,soubeyroux_ljpc_49_C8_1988spin} 
Early reports on the spin-glass phase in FeAl$_2$O$_4$ were based on experiments performed on samples with relatively significant degree of cation inversion (for example, 17 $\%$ reported in [\onlinecite{soubeyroux_ljpc_49_C8_1988spin}]). 
In FeAl$_2$O$_4$ samples with approximately 8 $\%$ cation inversion, proximity to spin-glass phase was suggested by Krimmel {\it et al}. But the reported quadratic $T$-dependence of specific heat of FeAl$_2$O$_4$ at low temperature is in contradiction with the $T$-linear response of canonical spin glasses. 
To explain this discrepancy, an argument based on spin-orbital liquid due to the presence of Jahn Teller Fe$^{2+}$ in tetrahedral-site was put forward in [\onlinecite{tristanprb_72_174404_2005geometric}] in comparison with observations on FeSc$_2$S$_4$  \cite{fritsch_prl_92_116401} and FeCr$_2$S$_4$. \cite{fichtl_prl_94_027601_2005}
\\
Our macroscopic magnetic measurements, ac susceptibility and {\it{memory effect}} test, have ruled out a spin glass scenario in FeAl$_2$O$_4$ samples with 
particularly low degree of cationic inversion. 
The apparent different behaviour of our samples (virtually no cation inversion and no spin-glass behaviour) and samples from earlier reports (large cation inversion, interpretation 
in the framework of spin-glass scenario) could be due to the additional disorder in the latter samples, which -- together with magnetic frustration -- could indeed lead to a 
spin-glass phase, so no unambiguous proof has been given in [\onlinecite{krimmelprb_79_134406_2009spin}] and [\onlinecite{tristanprb_72_174404_2005geometric}]. 
The diffuse magnetic scattering results presented here on FeAl$_2$O$_4$ are direct confirmation of the predominant short-range order present down to 4~K.
The analysis of magnetic scattered intensity shows that the spin-spin correlations in FeAl$_2$O$_4$ extend up to third-nearest-neighbour shells in the spinel unit cell. 
It is tempting to interpret the results on Fe, Mn-spinels based on the "order by disorder'' physics adopted to the case of $A$-site spinels which are diamond-lattice 
antiferromagnets. \cite{bergmannature_3_487_2007order} 
In this scenario, frustration is brought about by competition between the nearest and next-nearest-neighbour interactions, $J_1$ and $J_2$, and the magnetic ground state is tuned by the ratio of these values. 
Accordingly, in the weakly frustrated limit where 0 $\leq \left(\frac{J_2}{J_1}\right) \leq$ $\frac{1}{8}$, the magnetic ground state is the N\'{e}el phase with anti-parallel spins on the n.n's. 
\begin{table}
\caption{The observed ($F^2_{obs}$) and calculated ($F^2_{calc}$) intensities for the ($hkl$) reflections presented in Fig~\ref{fig_dns_mnal2o4_a} (c).}
\begin{tabular}{lllll} \hline\hline
($h$ $k$ $l$) & $F^2_{obs}$ & $F^2_{calc}$ & $d$-spacing \\ 
\hline\hline
(1   1   1)         &    0.195(2)    &     0.192    &  4.75287        \\            
(0   0   2)         &    0.912(2)    &     0.977    &  4.11611           \\         
(3   1   1)       &      0.034(4)     &    0.039     & 2.48211                    \\
(2   2   2)        &    0.199(3)      &    0.209    &  2.37644\\ \hline
\end{tabular}
\label{tab3}
\end{table}
As the value of $J_2$ increases, the N\'{e}el phase transforms to {\it spiral spin liquid} phase. Increase in frustration through increase in $J_2$ also results in the reduction of 
$T_N$ with respect to $\Theta_{\mathrm {CW}}$. 
In the case of FeAl$_2$O$_4$, with $\Theta_{\mathrm{CW}} \approx$ 121~K and $T_a \approx$ 13~K, a notable reduction in $T_a$ is clear which suggests the enhancement of $J_2$. 
The mean-field exchange parameter, $J_\mathrm{MF}$, reported \cite{krimmelprb_79_134406_2009spin} for FeAl$_2$O$_4$ is 16.25~K, which lies between those of 
MnAl$_2$O$_4$ ($J_\mathrm{MF}$ = 12.3~K) with $\left(\frac{J_2}{J_1}\right) \approx$ 0.09 and CoAl$_2$O$_4$ ($J_\mathrm{MF}$ = 20.8~K) with $\left(\frac{J_2}{J_1}\right) 
\approx$ 0.17. 
From these values, it is suggestive that the magnetic ground state evolves from N\'{e}el state in MnAl$_2$O$_4$ towards spin-liquid in FeAl$_2$O$_4$. 
This is reflected in the coexistence of long- and short-range order in Mn-spinel and the dominance of short-range spin fluctuations in the Fe-spinel in our study. 
In a detailed neutron scattering study on single crystal of CoAl$_2$O$_4$, MacDougall {\it et al.,} \cite{macdougall_pnas_108_15693_2011kinetically} argue that the phase 
transition observed at 6.5~K shows first-order nature with long-range magnetic order being inhibited by kinetic freezing of domain walls. 
While such a scenario based on kinetic freezing of domain walls can form the basis for glassy-like features observed in some of the $A$-site spinels, our magnetic measurements 
on FeAl$_2$O$_4$ show no signatures of glass-like freezing. 
From the present experimental results it is not possible to verify the presence, if at all, of clusters or domain-related liquid-like distributions which could be modeled by assuming 
rigid magnetic droplets.
\\
"Site inversion" wherein, the $A$-site and $B$-site cations exchange their respective crystallographic positions, is an important issue in spinels which has a strong bearing on their magnetic properties. 
Depending on the degree of inversion, quantified by the inversion parameter $x$, “normal ($x$ = 0)”, “random ($x$ = 2/3)” and “inverse ($x$ =1)” spinels exist. 
Through controlled synthesis the degree of inversion can be lowered\cite{tristanprb_72_174404_2005geometric} however, a finite amount of disorder exists in real compounds.
Previous works on spinels\cite{oneill_pcm_18_302_1991,harrison_am_83_1092_1998temperature} have reported the use of different temperatures of synthesis followed 
by different cooling rates and quenching/annealing procedures to prepare samples with different degrees of inversion.
\begin{figure}[!t]
\centering
\includegraphics[scale=0.26]{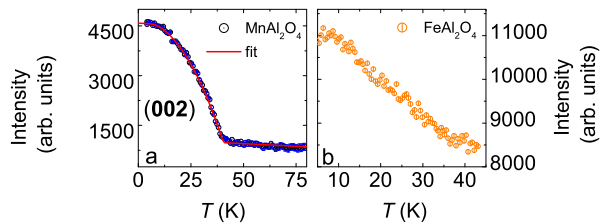}
\caption{(color online) (a) The temperature dependence of magnetic intensity of MnAl$_2$O$_4$, where the magnetic phase transition at $T_N \approx$ 40~K is clear. The solid red line shows the fit according to a mean-field model as per Eqn~(\ref{eqn_Bfit}). (b) In the case of FeAl$_2$O$_4$, a linear curve of dominant magnetic fluctuations is observed.}
\label{fig_magT_both}
\end{figure}
As mentioned in section $\ref{exp}$, the present set of samples were synthesized at 1000~$^\circ$C by employing slow rates of cooling in order to obtain low degree of inversion as suggested by the x-ray studies.
\\
The effect of site inverson on the spin liquid state of $A$-site spinels have been treated theoretically wherein the effect of weak disorder is found to act as a degeneracy breaking 
mechanism to select a preferred ground state out of the manifold of available states.\cite{savary_prb_84_064438_2011impurity}
The proposed "swiss cheese model" which takes in to account the disorder effects shows that the presence of finite amount of disorder does not diminish the frustration effects in an antiferromagnetic diamond lattice.
The samples of FeAl$_2$O$_4$ studied here contain low amounts of disorder, but even in the presence of it, frustration effects assume importance.
Recent work on CoAl$_2$O$_4$ by Roy {\it et al.,}\cite{roy_prb_88_174415_2013experimental} and Hanashima {\it et al.,}\cite{hanashima_jpsj_82_024702_2013spin} 
have taken into account the important parameter of site inversion to explain the magnetic properties of $A$-site spinels.
\\
In contrast to the case of FeAl$_2$O$_4$, MnAl$_2$O$_4$ is in the weakly frustrated regime with evidently an antiferromagnetic phase below $T_N$. 
As noted above, a speculative estimation of $\left(\frac{J_2}{J_1}\right)$ using mean-field approximation \cite{krimmelprb_79_134406_2009spin} and comparison with the 
theoretical calculations \cite{bergmannature_3_487_2007order} points towards N\'{e}el phase in MnAl$_2$O$_4$. From our polarized neutron measurements, sharp magnetic 
peaks are present at 4~K which shows that MnAl$_2$O$_4$ develops long-range order. 
Estimation of magnetic moment value through refinement of magnetic scattering data at 4~K leads to a value of 1.17(3)~$\mu_{\mathrm{B}}/{\mathrm{f.u.}}$. 
At 5~K, the macroscopically measured magnetic moment under applied field of 50~kOe was $\approx$ 0.4~$\mu_\mathrm{B}$/f.u. (see Fig~\ref{fig_mt_mnal2o4} (b)) which is 
diminished as compared to the spin-only ferromagnetic moment value of Mn$^{2+}$, 5~$\mu_\mathrm{B}$/f.u..  
Through the polarized neutron experiments we prove the existence of short-range order coexisting with long range ordered regions in the case of MnAl$_2$O$_4$; see 
Fig~\ref{fig_magT_both} and Fig~\ref{fig_dns_mnal2o4_a}. 
In this scenario, where short-range and long-range ordered phases coexist, a magnetic model of spin clusters or kinetically arrested domains suggested in the case of 
CoAl$_2$O$_4$ could also account for the features of diffuse magnetic scattering intensity.
\section{Conclusions}
To conclude, through polarized neutron scattering experiments and polarization analysis, we observe predominant short-range order in FeAl$_2$O$_4$, which extend 
up to third nearest-neighbour shells in spinel unit cell. 
The contribution from orbital degrees of freedom of Fe$^{2+}$ towards magnetism is suggested. 
Clear indication of long-range antiferromagnetic order at 4~K is obtained in the case of MnAl$_2$O$_4$.
However, significant short-range order coexist with long-range magnetic order in this compound even at $T_N \approx$ 40~K. 
A comparison of the magnetic behaviour of MnAl$_2$O$_4$ (Mn$^{2+}$; $S$ = $\frac{5}{2}$), FeAl$_2$O$_4$ (Fe$^{2+}$; $S$ = 2) and CoAl$_2$O$_4$ (Co$^{2+}$; $S$ = 
$\frac{3}{2}$) suggests a transition from magnetically long-range order to short-range order as the $S$ value is reduced. 
This provides a hint for investigating $S$ = 1 and $\frac{1}{2}$ spinels in search of quantum effects in spin liquid states. 
Through our re-investigation, we have convincingly shown that:
(i) FeAl$_2$O$_4$ displays no long-range magnetic order down to 4~K 
(ii) the true ground state of minimally disordered FeAl$_2$O$_4$ is not a canonical spin-glass
(iii) in MnAl$_2$O$_4$, long- and short-range spin correlations coexist. 
Remarkably, by considering quantum fluctuations as the predominant mechanism that relieves spin frustration, a phase diagram comprising of six coplanar spiral ordering in 
addition to N\'{e}el phase was suggested.\cite{bernier_prl_101_047201_2008quantum} 
A strong connection between frustration and strength of quantum fluctuations was suggested. 
However the important role played by the inversion parameter in choosing the magnetic ground state in $A$-site spinels cannot be discarded.
%
%
%
%

%
%
%
%
\end{document}